\documentclass[10pt,twocolumn,prl,aps,amssymb,amsmath,tightenlines,showpacs]{revtex4}

\newcommand{\cC}{\ensuremath{\mathcal{C}}}
\newcommand{\cP}{\ensuremath{\mathcal{P}}}
\newcommand{\cT}{\ensuremath{\mathcal{T}}}
\newcommand{\cQ}{\ensuremath{\mathcal{Q}}}
\newcommand{\half}{\mbox{$\textstyle{\frac{1}{2}}$}}

\begin{document}

\leftline{preprint LA-UR-07-3525}

\title{No-ghost theorem for the fourth-order derivative Pais-Uhlenbeck
oscillator model}

\author{Carl~M.~Bender${}^1$\footnote{Permanent address: Physics Department,
Washington University, St.~Louis, MO 63130, USA} and
Philip~D.~Mannheim${}^2$}

\affiliation{${}^1$Center for Nonlinear Studies, Los Alamos National
Laboratory,
Los Alamos, NM 87545, USA\\ {\tt electronic address: cmb@wustl.edu}\\ \\
${}^2$Department of Physics, University of Connecticut, Storrs, CT 06269, USA
\\{\tt electronic address: philip.mannheim@uconn.edu}}

\date{June 1, 2007}

\begin{abstract}
Contrary to common belief, it is shown that theories whose field equations are
higher than second order in derivatives need not be stricken with ghosts. In
particular, the prototypical fourth-order derivative Pais-Uhlenbeck oscillator
model is shown to be free of states of negative energy or negative norm. When
correctly formulated (as a $\cP\cT$ symmetric theory), the theory determines its
own Hilbert space and associated positive-definite inner product. In this
Hilbert space the model is found to be a fully acceptable quantum-mechanical
theory that exhibits unitary time evolution.
\end{abstract}


\maketitle

It is widely believed that field theories based on equations of motion higher
than second order are unacceptable. It is thought that higher-order theories
would possess propagators having poles with nonpositive residues. Such poles
would be associated with states, known as {\it ghosts}, which have nonpositive
norms and would therefore threaten the unitarity of the theory. The purpose of
this paper is to debunk this folklore, and in so doing, to regenerate interest
in higher-order quantum field theories. Higher-order field theories are
potentially of great value because they can address naturally the
renormalization issues connected with elementary particle self-energies and
quantum gravitational fluctuations \cite{r1}.

To understand the issues involved, we review the situation that arises with
regard to the Lee model. The Lee model was proposed in 1954 as a trilinearly coupled quantum
field theory in which the entire renormalization program can be carried out in
closed form \cite{r2}. However, just one year later it was argued that this
theory has a ghost state \cite{r3}. To be precise, a ghost appears to arise in
the Lee model when the renormalized coupling constant exceeds a critical value.
Above this critical value, the Lee-model Hamiltonian immediately becomes
non-Hermitian in the Dirac sense (Dirac Hermitian conjugation means combined
matrix transposition and complex conjugation) because its trilinear interaction term
acquires an imaginary coefficient. In the non-Hermitian phase of the Lee model a
state of negative norm in the Dirac sense emerges.

For the past half century, there have been multiple attempts to make sense of
the Lee model as a valid quantum theory (starting as early as \cite {r4}), but
it was not until 2005 that it was understood that if this theory is formulated
correctly, it does not have a ghost at all \cite{r5}. The solution to the
Lee-model-ghost problem is that when the coupling constant exceeds its critical
value, the Hamiltonian undergoes a transition from being Dirac Hermitian to
being $\cP\cT$ symmetric, that is, symmetric under combined parity reflection
and time reversal. In addition, in the sector of the Lee model in which the
ghost appears, the $\cP\cT$ symmetry is not broken in the sense that all energy
eigenvalues are real. For any $\cP\cT$-symmetric Hamiltonian having an unbroken
$\cP\cT$ symmetry, it is necessary to introduce a completely new Hilbert-space
inner product \cite{r6,r7,r8,r9,r10}. With respect to the new inner product
appropriate for the $\cP\cT$-symmetric phase of the Lee model, the Hamiltonian
becomes self-adjoint, and its so-called ghost state becomes an ordinary quantum
state with positive $\cP\cT$ norm. This same procedure has been applied to other
seemingly problematic models \cite{r11}.

The purpose of this paper is to show that this same prescription not only can,
but in fact must be implemented in higher-derivative field theories as well, and
that the negative Dirac-norm states that arise in these theories can instead
really be ordinary quantum states having positive $\cP\cT$ norm. Specifically,
we will use the same technique developed for the Lee model to study the
fourth-order quantum-mechanical Pais-Uhlenbeck oscillator, a model which is the
prototypical higher-derivative quantum field theory. We will show by explicit
construction that this model is actually a $\cP\cT$ symmetric theory that is
totally free of negative-norm states.

The action that defines the Pais-Uhlenbeck model is acceleration-dependent
\cite{r12}:
\begin{equation}
I_{\rm PU}=\frac{\gamma}{2}\int dt\left[{\ddot z}^2-\left(\omega_1^2
+\omega_2^2\right){\dot z}^2+\omega_1^2\omega_2^2z^2\right],
\label{e1}
\end{equation}
where $\gamma$, $\omega_1$, and $\omega_2$ are all positive constants and 
without loss of generality we take $\omega_1\geq\omega_2$. This model represents
two oscillators coupled by a fourth-order equation of motion of the form
$d^4z/dt^4+(\omega_1^2+\omega_2^2)d^2z/dt^2+\omega_1^2\omega_2^2z=0$
\cite{r13}. With ${\dot z}$ serving as the canonical conjugate of both
$z$ and ${\ddot z}$, the system is constrained and its Hamiltonian must be
found by the method of Dirac constraints. To this end, in place of ${\dot 
z}$ we introduce a new dynamical variable $x$ (with corresponding conjugate $p_x$), and via the Dirac method construct the Hamiltonian \cite{r14,r15,r16}
\begin{equation}
H=\frac{p_x^2}{2\gamma}+p_zx+\frac{\gamma}{2}\left(\omega_1^2+\omega_2^2
\right)x^2-\frac{\gamma}{2}\omega_1^2\omega_2^2z^2.
\label{e2}
\end{equation}
This Hamiltonian depends on two coordinates $x$ and $z$, and their canonical
conjugates, $p_x$ and $p_z$. Using this Hamiltonian, the Poisson-bracket algebra
of the operators $x$, $p_x$, $z$, and $p_z$ is found to be closed with its
nonzero elements given by $\{x,p_x\}=1$ and $\{z,p_z\}=1$. This construction
makes no reference to the classical equations of motion and thus holds for both
stationary and nonstationary classical paths. Consequently, we can use it to
quantize the theory, with the nonzero quantum commutators being given by
$[x,p_x]=i$ and $[z,p_z]=i$.

To construct a Fock-space representation of the theory, we introduce two sets of
creation and annihilation operators according to
\begin{eqnarray}
z&=&a_1+a_1^\dagger+a_2+a_2^\dagger,
\nonumber \\
p_z&=&i\gamma\omega_1\omega_2^2
(a_1-a_1^\dagger)+i\gamma\omega_1^2\omega_2(a_2-a_2^\dagger),\nonumber\\
x&=&-i\omega_1(a_1-a_1^\dagger)-i\omega_2(a_2-a_2^\dagger),\nonumber\\
p_x&=&-\gamma\omega_1^2 (a_1+a_1^\dagger)-\gamma\omega_2^2(a_2+a_2^\dagger),
\label{e3}
\end{eqnarray}
This yields a Hamiltonian of the form \cite{r14,r15,r16}
\begin{equation}
H=2\gamma(\omega_1^2-\omega_2^2)(\omega_1^2 a_1^\dag a_1-\omega_2^2a_2^\dag a_2)
+\half(\omega_1+\omega_2),
\label{e4}
\end{equation}
with the nonzero Fock-space commutators being given by
\begin{equation}
\omega_1[a_1,a_1^\dag]=-\omega_2[a_2,a_2^\dag]=\frac{1}{2\gamma
\left(\omega_1^2-\omega_2^2\right)}.
\label{e5}
\end{equation}

The Hamiltonian (\ref{e4}) describes two harmonic oscillators, each with
strictly real energy eigenvalues. There are two possible realizations for the
theory. If we take $a_1$ and $a_2$ to annihilate the no-particle state $|\Omega
\rangle$,
\begin{equation}
a_1|\Omega\rangle=0, \qquad a_2|\Omega\rangle=0,
\label{e6}
\end{equation}
the energy spectrum is then bounded below, and in this case $|\Omega\rangle$ is
the ground state with energy $\half\left(\omega_1+\omega_2\right)$.
However in this case, the excited state $a_2^\dag|\Omega\rangle$, which lies at
energy $\omega_2$ above the ground state, has a Dirac norm $\langle\Omega
|a_2a_2^\dag|\Omega\rangle$ which is negative. On the other hand, if $a_1$ and $a_2^\dag$
annihilate the no-particle state $|\Omega\rangle$,
\begin{equation}
a_1|\Omega\rangle=0,\qquad a_2^\dag|\Omega\rangle=0,
\label{e7}
\end{equation}
then the theory is free of negative-norm states, but the energy spectrum is
unbounded below. Both of these outcomes are unacceptable and characterize the
generic problems that are thought to afflict higher-derivative quantum theories.

The above analysis relies on using the standard Dirac norm, but such a norm in
quantum theory is actually not mandatory. Specifically, in an eigenvalue
equation of the form $H|\psi\rangle=E|\psi\rangle$, the Hamiltonian acts
linearly on ket vectors, without any reference to bra vectors at all.
Consequently, the eigenvalue equation is unaffected if the bra vector is not
taken to be the Dirac conjugate $\langle\psi|$ of the ket $|\psi\rangle$.

To determine what the appropriate bra vector should be, we need to supply some
global information. This is best done using the coordinate representation of the
canonical commutators, which for the Pais-Uhlenbeck oscillator model are given
by
\begin{equation}
p_z=-i\frac{\partial}{\partial z},\qquad p_x=-i\frac{\partial}{\partial x}.
\label{e8}
\end{equation}
In this representation the Schr\"odinger equation $H\psi_n=E_n\psi_n$ takes the
form
\begin{eqnarray}
&&\bigg[-\frac{1}{2\gamma}\frac{\partial^2}{\partial x^2}
-ix\frac{\partial}{\partial z}+\frac{\gamma}{2}\left(\omega_1^2+\omega_2^2
\right)x^2\nonumber\\
&&\qquad-\frac{\gamma}{2}\omega_1^2\omega_2^2z^2\bigg]\psi_n(z,x)=
E_n\psi_n(z,x),
\label{e9}
\end{eqnarray}
the state whose energy is $\half\left(\omega_1+\omega_2\right)$ has an
eigenfunction of the form \cite{r16}
\begin{eqnarray}
\psi_0(z,x)&=&\exp\bigg[\frac{\gamma}{2}(\omega_1+\omega_2)
\omega_1\omega_2z^2
\nonumber\\
&&\qquad+i\gamma\omega_1\omega_2zx -\frac{\gamma}{2}(\omega_1+\omega_2)x^2
\bigg],
\label{e10}
\end{eqnarray}
and the states of higher energy have eigenfunctions that are polynomial
functions of $x$ and $z$ times $\psi_0(z,x)$. The eigenfunction $\psi_0(z,x)$
has the defect that it is not normalizable on the real-$x$ and real-$z$ axes; it
grows exponentialy as $z\to\pm\infty$. Evidently, the coordinate-space
realization of $p_z$ in (\ref{e8}) is not Hermitian.

Since one can only use the realization $p_z=-i\partial/\partial z$ when the $[z,p_z]
$ commutator acts on well-behaved test functions, we see that such well-behaved
functions cannot be taken to lie on the real $z$ axis. Moreover, the divergence
of $\psi_0(z,x)$ in (\ref{e10}) as $|z|\to\infty$ is not restricted to the real
axis, and it even occurs in two {\it Stokes's wedges} in the complex-$z$ plane
of angular opening $90^\circ$ and centered about the positive- and negative-real
axes (the east and west quadrants of the letter $X$). However, in the
complementary $90^\circ$ Stokes' wedges centered about the positive- and
negative-imaginary axes (the north and south quadrants of the letter $X$),
$\psi_0(z,x)$ vanishes exponentially rapidly as $|z|\to\infty$. We must thus
restrict the eigenvalue problem (\ref{e9}) to the complementary (north, south)
Stokes' wedges, and in so doing we thus take care of the normalization problem
for the eigenfunctions. In these wedges $\psi_0(z,x)$ is the fully normalizable
ground-state of the system and the energy spectrum is precisely the purely real
one associated with the Fock-space option given in (\ref{e6}).

Rather than working on the imaginary axis, it is instructive to perform the
$90^\circ$ rotation $y=-iz$. The modified Hamiltonian with $y=-iz$ (and
therefore with $q=ip_z$ to enforce $[y,q]=i$) has the form
\begin{equation}
H=\frac{p^2}{2\gamma}-iqx+\frac{\gamma}{2}\left(\omega_1^2+\omega_2^2
\right)x^2+\frac{\gamma}{2}\omega_1^2\omega_2^2y^2,
\label{e11}
\end{equation}
where for notational simplicity we have replaced $p_x$ by $p$. In (\ref{e11})
the operators $p$, $x$, $q$, and $y$ are now formally Hermitian \cite{r17}, but
because of the $-iqx$ term, $H$ has become complex and is manifestly not Dirac
Hermitian \cite{r18}. This non-Hermiticity property is not at all apparent in
the original form of the Hamiltonian given in (\ref{e2}). This surprising and
completely unexpected emergence of a non-Hermitian term in the Pais-Uhlenbeck
Hamiltonian is the root cause of the infamous ghost problem of the
Pais-Uhlenbeck model.

While the Hamiltonian in (\ref{e11}) is not Dirac Hermitian, it does fall into a
particular class of equally physically viable Hamiltonians, those that are $\cP
\cT$ symmetric. To establish this $\cP\cT$ symmetry we make the following
assignments: Under $\cP$ and $\cT$, we take $p$ and $x$ to transform like
conventional coordinate and momentum variables. However, we define $q$ and $y$
to transform unconventionally in a way that has not been seen in previous
studies of $\cP\cT$-symmetric quantum mechanics; in the language of quantum
field theory, $q$ and $y$ transform as parity scalars instead of pseudoscalars, and
they have abnormal behavior under time reversal. These transformation
properties are summarized in the following table:
\begin{equation}
\begin{array}{c|cccc}
&p&x&q&y\\
\hline\\
\cP&-&-&+&+\\
\cT&-&+&+&-\\
\cP\cT&+&-&+&-\\
\end{array}
\label{e12}
\end{equation}
Because $H$ has an entirely real spectrum, the $\cP\cT$ symmetry of $H$ is
unbroken.

Having shown that the $\cP\cT$ symmetry of $H$ is unbroken, we will be able
to reinterpret the ghost as a conventional quantum state of positive $\cP\cT$
norm. Specifically, following the standard procedures of $\cP\cT$-symmetric
quantum mechanics, we construct the $\cP\cT$ norm by calculating an operator
called the $\cC$ operator \cite{r19}. The $\cC$ operator associated with the
Hamiltonian of interest in (\ref{e11}) is required to satisfy a characteristic
set of three conditions:
\begin{equation}
\cC^2=1,\quad[\cC,\cP\cT]=0,\quad [\cC,H]=0.
\label{e13}
\end{equation}
This first two of these conditions are kinematical, while the third is dynamical
since it involves the specific Hamiltonian $H$.

In previous investigations it has been established that $\cC$ has the general
form $\cC=e^{-\cQ}\cP$, where $\cQ$ is a real function of the dynamical
variables and is Hermitian in the Dirac sense, and it was found that $\cQ$ was
odd under a change in sign of the momentum variables and even under a change in
sign of the coordinate variables. However, because of the abnormal behaviors of
the $y$ and $q$ operators in (\ref{e12}), we find that the exact solution to the
three simultaneous algebraic equations in (\ref{e13}) gives an unusual and
previously unencountered structure for $\cQ$:
\begin{equation}
\cQ=\alpha pq+\beta xy,
\label{e14}
\end{equation}
where $\alpha$ and $\beta$ are given by
\begin{eqnarray}
\beta&=&\gamma^2\omega_1^2\omega_2^2\alpha,\quad
\sinh(\sqrt{\alpha\beta})=\frac{2\omega_1\omega_2}{(\omega_1^2-\omega_2^2)}.
\label{e15}
\end{eqnarray}

Even though the form of $\cQ$ in (\ref{e14}) is unprecedented in $\cP
\cT$-symmetric quantum mechanics, the effect of performing a similarity
transformation on the dynamical variables $x,p,y,q$ using $e^{-\cQ}$ still
generates a canonical transformation that preserves the commutation relations,
as has always been the case in the past. For the Pais-Uhlenbeck Hamiltonian, the
transformation is \cite{r20}:
\begin{eqnarray}
e^{-\cQ}xe^\cQ&=&x\cosh(\sqrt{\alpha\beta})+i\sqrt{\frac{\alpha}{\beta}}\,q\sinh
(\sqrt{\alpha\beta}),\nonumber\\
e^{-\cQ}qe^\cQ&=&q\cosh(\sqrt{\alpha\beta})-i\sqrt{\frac{\beta}{\alpha}}\,x\sinh
(\sqrt{\alpha\beta}),\nonumber\\
e^{-\cQ}ye^\cQ&=&y\cosh(\sqrt{\alpha\beta})+i\sqrt{\frac{\alpha}{\beta}}\,p\sinh
(\sqrt{\alpha\beta}),\nonumber\\
e^{-\cQ}pe^\cQ&=&p\cosh(\sqrt{\alpha\beta})-i\sqrt{\frac{\beta}{\alpha}}\,y\sinh
(\sqrt{\alpha\beta}).
\label{e16}
\end{eqnarray}

In $\cP\cT$-symmetric quantum mechanics, performing a similarity transformation
on the $\cP\cT$-symmetric Hamiltonian with $e^{-\cQ/2}$ yields a positive
definite Hamiltonian which is Hermitian in the Dirac sense \cite{r21}. In the 
current context we obtain
\begin{eqnarray}
{\tilde H}&=&e^{-\cQ/2}He^{\cQ/2}\nonumber\\
&=&\frac{p^2}{2\gamma}+\frac{q^2}{2\gamma\omega_1^2}+
\frac{\gamma}{2}\omega_1^2x^2+\frac{\gamma}{2}\omega_1^2\omega_2^2y^2.
\label{e17}
\end{eqnarray}
The spectrum of this Hamiltonian is manifestly real and positive. However,
because this Hamiltonian is related to the original Pais-Uhlenbeck Hamiltonian
by a similarity transformation, which is isospectral, despite the $-iqx$ term,
the positivity of the Pais-Uhlenbeck Hamiltonian is proved.

Furthermore, the eigenstates of $|{\tilde \psi}\rangle$ of ${\tilde H}$ have
positive inner product and can be normalized in the conventional Dirac way using
the standard inner product:
\begin{equation}
\langle{\tilde \psi}|{\tilde\psi}\rangle=1,
\label{e18}
\end{equation}
where the bra vector is the Dirac-Hermitian adjoint of the ket vector.
Equivalently, for the eigenstates $|\psi\rangle$ of the Hamiltonian $H$, because
the vectors are mapped by
\begin{equation}
|{\tilde \psi}\rangle=e^{-\cQ/2}|\psi\rangle,
\label{e19}
\end{equation}
the normalization of the eigenstates of $H$ is given as
\begin{equation}
\langle\psi|e^{-\cQ}|\psi\rangle=1.
\label{e20}
\end{equation}
Thus, it is the norm of (\ref{e20}) that is relevant for the Pais-Uhlenbeck
model, with $\langle\psi | e^{-\cQ}$ rather than $\langle \psi|$ being the
appropriate conjugate for $|\psi\rangle$. Finally, because the norm in
(\ref{e20}) is positive and because $[H,\cC\cP\cT]=0$, the Hamiltonian $H$
generates unitary time evolution.

To conclude, we see that in order to construct the correct Hilbert space for the
Pais-Uhlenbeck theory, one must determine the region in the complex plane where
operators such as $-i\partial/\partial_z$ act as well-defined Hermitian
operators. The appearance of a ghost state when one takes the derivative
operator to be Hermitian on the real $z$ axis is not an indication that there is
anything wrong with the theory itself, but only with the way it is being
analysed. Consequently, if treated properly, higher derivative theories such as
conformal gravity have the potential to be completely viable as quantum theories
of gravity in four spacetime dimensions \cite{r22}.

As an Ulam Scholar, CMB receives financial support from the Center for
Nonlinear Studies at the Los Alamos National Laboratory. CMB is also supported
by a grant from the U.S. Department of Energy.

{}

\begin{thebibliography}{}
\bibitem{r13} With the identifications $\omega_1^2+\omega_2^2=2k^2+M^2$, 
$\omega_1^2\omega_2^2=k^4+k^2M^2$, this equation of motion is
the quantum-mechanical limit of the field-theoretic second-plus-fourth order
$(-\partial_t^2+\nabla^2)(-\partial_t^2+\nabla^2-M^2)\phi(\bar{x},t)=0$ in
field configurations of the form $\phi(\bar{x},t)=z(t)e^{i\bar{k}\cdot\bar{x}}$. The $M=0$ pure fourth order theory corresponds to $\omega_1=\omega_2=k$.
\bibitem{r17} Despite the introduction of $y=-iz$ and $q=ip_z$, since $y$
and $q$ are Hermitian operators, their c-number eigenvalue coordinates and
momenta are strictly real.
\bibitem{r18} Although quantum mechanics is usually formulated in terms of
Hermitian operators, Hermiticity is only sufficient to give real eigenvalues but
not necessary. Specifically, while a Hermitian operator has real eigenvalues,
there is no converse theorem that says that the eigenvalues of a non-Hermitian
operator are not real. The Hamiltonian (\ref{e11}) is an example of a
non-Hermitian operator that possesses a strictly real spectrum of eigenvalues.
\bibitem{r22} In \cite{r14,r15,r16} it was shown that even if one uses the Dirac inner
product for the Pais-Uhlenbeck theory, in the equal frequency limit (the case
relevant to conformal gravity) the states of negative norm decouple from the
Hamiltonian. With the Hamiltonian of (\ref{e17}) being completely well-behaved
in the equal frequency limit, we confirm the viability of the equal-frequency
limit of the theory.
\end{thebibliography}
\end{document}